\documentclass[twocolumn]{aastex631} 
\usepackage[switch,mathlines]{lineno}
\usepackage{graphicx} 
\usepackage{amsmath}   
\usepackage{booktabs} 
\usepackage{placeins}

\begin{document}

\flushbottom
\title{Elemental Abundances in the Binary Star V505 Per}
\author{Zohreh Safari Forushani }
\affiliation{Department of Astronomy, Indiana University, Bloomington, IN 47405, USA}
\author{Catherine A. Pilachowski}
\affiliation{Department of Astronomy, Indiana University, Bloomington, IN 47405, USA}
\author{Gloria Koenigsberger}
\affiliation{Instituto de Ciencias Físicas, Universidad Nacional Autónoma de México, México}
\author{Derek Sikorski}
\affiliation{Institute for Astronomy, University of Hawai‘i, 2680 Woodlawn Drive, Honolulu, HI 96822, USA}
\author{Maria Cordero}
\affiliation{Department of Astronomy, Indiana University, Bloomington, IN 47405, USA}

\begin{abstract}
We present a detailed chemical abundance analysis of the eclipsing binary system V505 Per. High-resolution spectra were analyzed using the \texttt{MOOG} spectrum analysis code, we determined abundances not only for iron and lithium but also for Si, Na, Ca, Mn, and Ni — elements that have not previously been analyzed in detail for this system. Abundances were computed across 15 temperature points using model atmospheres, with stellar parameters refined by minimizing abundance trends with excitation potential. We determined effective temperatures of \(T_{\mathrm{eff}} = 6650\pm 50~\mathrm{K}\) for the primary and \(T_{\mathrm{eff}} = 6550\pm 50~\mathrm{K}\), with iron abundances of [Fe/H] $= -0.10 \pm 0.06$ and [Fe/H] $= -0.19 \pm 0.07$, respectively. Most [X/Fe] ratios are consistent with solar values, though manganese is deficient. Our analysis of the effective temperatures shows that both stars lie on the hot edge of the lithium dip, consistent with \citet{2025arXiv250719954K}, which may help resolve the inconsistency noted of the stars lithium abundance within the dip by \citet{2013PASP..125..753B}.


\end{abstract}

\section{Introduction}
V505 Per( HD 14384; SAO 23229) is an eclipsing, double-lined spectroscopic binary system with $V=6.88\pm0.01$ \citep{2021Obs...141..234S} located in the constellation Perseus. It was first identified by \citep{1989JAVSO..18..149K} who identified periodic eclipses with an amplitude of approximately 0.5 magnitudes. The system consists of two nearly identical F5~V main-sequence stars. 
 
 Later spectroscopic observations confirmed the system to be double-lined, with nearly equal-mass components on a circular orbit. The components orbit each other with a period of 4.22 days \citep{1997AJ....114..793M}. Their spectral classification suggests surface temperatures in the range of approximately  $\sim$6400–6500 K, consistent with moderately young solar-type stars.

Orbital phase information provides the basis for the  precise identification of eclipse events and the monitoring of photometric variability in the V505 Per system \citep{2021Obs...141..234S}. This binary consists of components A and B, with component A being slightly brighter, larger, and more massive than component B. The masses of stars are $M_A = 1.2745\,M_\odot$ and $M_B = 1.2577\,M_\odot$, and their radii are $R_A= 1.2941\,R_\odot$, and $R_B= 1.2637\,R_\odot$ \citep{2021Obs...141..234S}. In orbital phase $ \phi = 0 $, component A is eclipsed by component B, corresponding to the primary eclipse. At $ \phi = 0.5 $, component B is eclipsed by component A, marking the secondary eclipse.

While the masses, radii, and temperature difference of the two components of V505 Per are well constrained by the light curve and orbital parameters of the binary, the effective temperatures of the two stars are less constrained. \citet{2008A&A...480..465T} found temperatures of $T_{\mathrm{eff}} = 6512\,\mathrm{K}$ and $6460\,\mathrm{K}$, while \citet{2025arXiv250719954K} suggested somewhat hotter temperatures of $T_{\mathrm{eff}} = 6600\,\mathrm{K}$ and $6550\,\mathrm{K}$.

The modest discrepancy in temperature between \citet{2008A&A...480..465T} and \citet{2025arXiv250719954K} becomes significant when considering  the lithium dip \cite{boesgaard1986lithium}. \citet{2013PASP..125..753B} adopted the \citet{2008A&A...480..465T} temperatures for V505 Per, which place the stars in the center of the lithium dip.  However, they noted that the lithium abundance is higher than expected for stars of similar age at these temperatures, and suggested that revisiting the effective temperatures may help resolve this discrepancy. Alternatively, they suggest that rapid tidal synchronization could have reduced the degree of lithium destruction compared to otherwise normal stars. At the higher temperatures suggested by \citet{2025arXiv250719954K}, V505~Per instead lies on the hot edge of the dip and its lithium abundances are consistent with normal stars of similar age and temperature. The purpose of our study is to determine which temperature scale is correct.  In addition, we aim to provide abundances of a larger set of elements than previously reported (\citet{2008A&A...480..465T}, \citet{2013PASP..125..753B}, and \citet{2025arXiv250719954K}.

 In the present work, we extend earlier efforts by measuring a larger set of  Fe absorption lines and including Li, Si, Na, Ca, Mn and Ni which allows us to provide a more complete chemical profile of the system.

 We report an equivalent width (EW) analysis to determine stellar parameters and elemental abundances. Section~2 describes the observations, data reduction procedures. Section~3 details the equivalent width measurements. In Section~4, we discuss the determination of stellar atmospheric parameters. Section~5 presents the chemical abundance analysis for both components of V505~Per. Specifically, we examine the abundances of Fe\,\textsc{i}, Na\,\textsc{i}, Si\,\textsc{i}, Ca\,\textsc{i}, Mn\,\textsc{i}, Ni\,\textsc{i}, and lithium. Section 6 explains the uncertainties in our abundance measurements, including random errors from the data and how sensitive the results are to the chosen stellar parameters. Section~7 discusses the results, and Section~8 summarizes our conclusions.

\section{Observations and Data Reduction}
Observations were obtained on UT date 2011-09-21, at Kitt Peak National Observatory Coudé Feed telescope using the Coudé Feed spectrograph. The instrumental setup included an echelle grating, a 250 $\mu m$ slit width, Camera~5, and an F3KB CCD detector. This configuration yielded a reciprocal dispersion of 1.9\,$\AA~ mm ^{-1}$ $(0.02 ~\AA~pixel^{-1})$ with a CCD pixel size of $0.015~mm$, providing a spectral resolving power of $R \approx 110{,}000$ over the observed wavelength range.  The components of V505 Per have modest projected rotational velocities, $v sin i \approx12.5  \pm 0.5 ~km s^{-1}$ \citep{2025arXiv250719954K}, comparable to the instrumental resolution and sufficient to ensure well-resolved spectral lines in both stars.

We obtained a 3600 s exposure of V505 Per, giving S/N ratios per pixel of about 119 at 5600~\AA ~and 85 at 6700~\AA.
The Julian Date of the observation was 2455825.98713, corresponding to an orbital phase of $\phi=0.9464$, based on an initial epoch of $T_0 = 2458798.516720$ and an orbital period of $P = 4.22201933$~days \citep{2025arXiv250719954K}. 
Spectral reduction was performed using the IRAF Echelle reduction package for spectral extraction and calibration, \citep{tody1986iraf,tody1993iraf,valdes1992iraf}.

\begin{figure*}[t]
    \centering
    \includegraphics[width=0.9\textwidth]{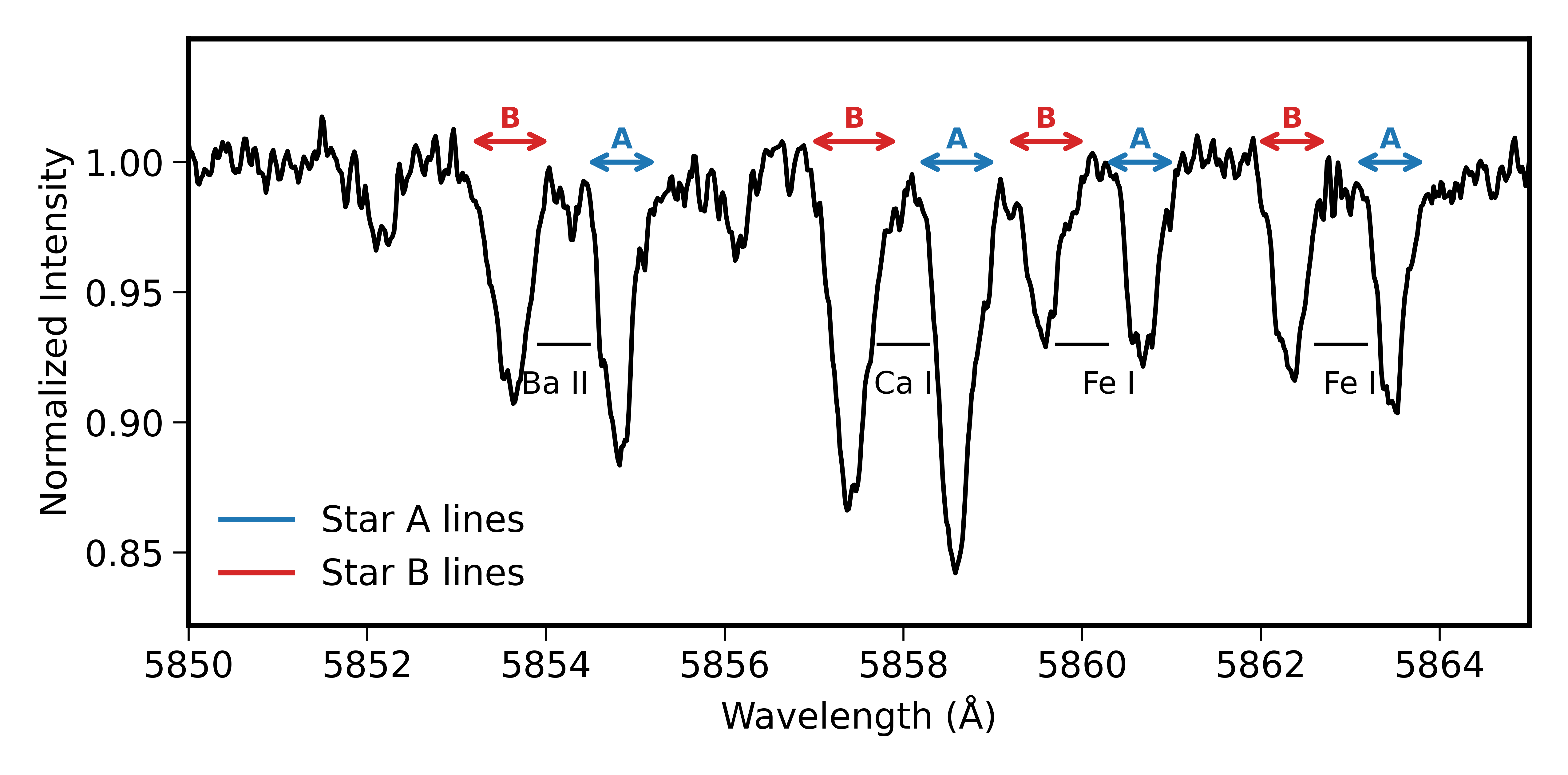}
    \caption{Illustrative segment of the  normalized  spectrum of V505 Per. \vspace{3em}}
    \label{fig:v505per_Fe}
\end{figure*}

To facilitate line identification, the absorption lines of the cooler component (star B) were aligned with laboratory wavelengths. Consequently, the spectral lines of the hotter component (star A) appear Doppler-shifted to the red relative to star B, corresponding to a relative shift of +58~km~s$^{-1}$ \citep{2025arXiv250719954K}segment with several identified lines is shown in Figure~\ref{fig:v505per_Fe}.\\

\section{Equivalent Width Analysis}
\subsection{Resolution Comparison Using the Solar Spectrum}
We compared our spectra with the solar spectrum to identify usable lines, using the disk–center solar spectrum from the \citet{Wallace1998} atlas. Here, usable lines are those that are unblended, reliably measurable, and sufficiently weak to avoid strong saturation effects.

To minimize saturation effects and reduce sensitivity to microturbulence velocity, we rejected lines with reduced equivalent width
$\log(W/\lambda) > -4.70$,
where $W$ is the equivalent width and $\lambda$ is the line wavelength (in the same units).
The same reduced-EW cutoff was applied to all species, including the Fe~I lines used to determine \(\xi_{\rm turb}\) and \(T_{\mathrm{eff}}\).

We identified 76 high-quality, unblended lines for component A and 71 for component B, including 56 Fe I lines free of blending. The equivalent widths of these lines were measured using Gaussian profile fitting with the splot task in IRAF. Lines of each component were also compared to the solar spectrum to ensure correct identification and to avoid blended lines.

Uncertainties in the equivalent widths were estimated by slightly shifting the continuum level up and down and re-measuring each line; the resulting change in EW was taken as the measurement uncertainty, typically about 2~m\AA.

\subsection{Equivalent width correction}

To determine the chemical abundances of the components in V505~Per, we must account for the relative contributions of each star to the total observed flux. Since both components are visible in the combined spectrum, the measured equivalent widths must be corrected using continuum weighting factors. The continuum weighting factors \(w_A\) and \(w_B\) represent the fractional continuum light contributions of components A and B, with the constraint \(w_A + w_B = 1\).

Following the method described by \citet{2008A&A...480..465T}, the effective temperatures of stars A and B were estimated using blackbody spectral energy distributions, assuming that both stars radiate approximately as blackbodies in the visual region. Additionally, following the approach of \citet{2025arXiv250719954K},   the relative weights used in the analysis were derived from the stellar luminosities, which depend on the radii of the stars ($R_{\mathrm{B}} / R_{\mathrm{A}} = 0.97895$) and the effective temperatures ($T_{\mathrm{B}} / T_{\mathrm{A}} = 0.9928$).

Assuming blackbody radiation in the optical regime, the ratio of continuum contributions can be expressed as:

\begin{equation}
\frac{w_A}{w_B} = \left( \frac{R_A}{R_B} \right)^2 \left( \frac{T_A}{T_B} \right)^4
\end{equation}
\hspace{0.5em}

The weighting factors  $w_A$ and  $w_B$ were adopted from \citet{2025arXiv250719954K}, who derived these values from the continuum flux ratio obtained through their curve-of-growth analysis and obtained $w_A / w_B = 1.074$, which yields $w_A = 0.518$ and $w_B = 0.482$ under the constraint $w_A + w_B = 1$. These weights were adopted and applied to correct the measured equivalent width for the relative light contribution of each stellar component.

These weights remain approximately constant across the wavelength range analyzed. Furthermore, varying the effective temperatures adopted by $\pm 100$~K (from 6512~K to 6600~K) changes the flux weighting factors by only $\sim 0.03\%$, indicating that the derived weights are independent of small uncertainties in $T_{\rm eff}$.Thus, the continuum flux ratio is effectively temperature-independent at our precision. The corrected equivalent widths were then used in the abundance calculations and are included in Table 1.
\hspace{0.5em}

\section{Stellar Parameters}

In order to determine the chemical abundances of both stellar components, we first re-examined their fundamental parameters, including effective temperature (\(T_{\mathrm{eff}}\)). 

\subsection{Effective Temperature}
We estimated $T_{\rm eff}$ for each star by minimizing the absolute value of the slope of \([\mathrm{Fe/H}]\) versus excitation potential (EP) for Fe\,\textsc{I} lines, over the temperature range \(6250\)–\(6950\) K in steps of 50 K. 
The slope is closest to zero at \(T_{\rm eff}=6650\pm50\) K for component A and \(6550\pm50\) K for component B.  
These temperatures are consistent, within the uncertainties, with the values \(6600\pm50\) K and \(6550\pm50\) K reported from line-profile fitting \citep{2025arXiv250719954K,2008A&A...480..465T}.
\vspace{0.5cm}

\subsection{Microturbulent Velocity}
We determined the microturbulent velocity, \(\xi_{\rm turb}\), by requiring no trend of \([\mathrm{Fe/H}]\) with reduced width, \(\log(EW/\lambda)\).  
Following \citet{2013PASP..125..753B}, we adopted  the calibration of \citet{prieto2004sn} for an initial estimate of the microturbulence of \(\xi_{\rm turb}=1.73\ \mathrm{km\,s^{-1}}\). We explored trial values of 1.00, 1.73, 2.00, and 2.30 \(\mathrm{km\,s^{-1}}\).  For each trial value we computed the slope of \([\mathrm{Fe/H}]\) versus \(\log(EW/\lambda)\) and adopted \(\xi_{\rm turb}=2.00\ \mathrm{km\,s^{-1}}\) as the value that gives the flattest trend, that is, a slope closest to zero. 
The quoted uncertainty of \(\pm0.20\ \mathrm{km\,s^{-1}}\) corresponds to the range of \(\xi_{\rm turb}\) for which the slope remains statistically consistent with zero.

\subsection{Surface Gravity}
The surface gravity (\(\log g\)=4.3) of each component was derived from the orbital elements (masses and radii) adopted from \citet{2008A&A...480..465T}.  
Varying \(\log g\) within the adopted uncertainties has a negligible effect on the derived iron abundances, as discussed in Section 6.

\setlength{\tabcolsep}{5.0pt}      
\renewcommand{\arraystretch}{1.5} 
\begin{deluxetable*}{cccccccccccc}
\tablecaption{Fe I measurement in V505 Per\label{tab:FeI}}
\tabletypesize{\scriptsize}
\tablehead{
  \multicolumn{6}{c}{\textbf{}} &
  \multicolumn{6}{c}{\textbf{}} \\
  \colhead{WL($\AA$)} & \colhead{$EW_{SUN}$($m\AA$)} & \colhead{EW$_A$($m\AA$)} & \colhead{EW$_B$($m\AA$)} & \colhead{[Fe/H]$_A$} & {[Fe/H]$_B$}& \colhead{WL($\AA$)} & \colhead{$EW_{SUN}$} & \colhead{EW$_A$($m\AA$)} & \colhead{EW$_B$($m\AA$)} & \colhead{[Fe/H]$_A$} & \colhead{[Fe/H]$_B$}
}
\startdata
5432.948&72& \ldots&66& \ldots&0.10 & 5987.065&71&34& 23&-0.44& \ldots\\
5445.042&118&\ldots&96& \ldots&-0.01 & 6003.012&85& \ldots&73& \ldots&0.10\\
5473.901&79&42& \ldots&-0.38& \ldots& 6007.960&62&55&41&0.08&-0.20\\
5487.745&89&81&80&-0.03&-0.03 &6008.557&88&58& \ldots&-0.35& \ldots\\
5543.147&62&\ldots&28&\ldots&-0.28 & 6024.058&110&72&76&-0.24&-0.26\\
5546.506&52&28& \ldots&-0.16& \ldots & 6027.051&64&36&32&-0.28&-0.41\\
5553.577&62& \ldots&24& \ldots&-0.51 & 6056.005&75&51&38&-0.11&-0.38\\
5554.895&98&58& \ldots&-0.31& \ldots & 6065.482&118&99& \ldots&0.07& \ldots\\
5557.983&65&27& \ldots&-0.42& \ldots & 6127.906&48&34&\ldots&0.01&\ldots\\
5560.212&53&36&23&-0.02&-0.32& 6157.728&62& \ldots&46& \ldots&-0.27\\
5562.706&62& \ldots&47& \ldots&-0.12 & 6173.334&67& \ldots&34& \ldots&-0.27\\
5563.600&102&56& \ldots&-0.45& \ldots& 6187.989&47&18& \ldots&-0.21& \ldots\\
5565.704&89& \ldots&66& \ldots&-0.24& 6200.313&73&44&46&-0.16&-0.20\\
5567.274&62&40& \ldots&-0.19& \ldots& 6213.430&82& \ldots&56& \ldots&-0.20\\
5572.842&258& \ldots&139& \ldots&-0.16& 6215.143&71&71& \ldots&0.17& \ldots\\
5576.089&117&86&780&-0.02&-0.21 & 6219.280&89&78&59&0.12&-0.26\\
5584.773&37&16&15&-0.11&-0.21&  6230.722&148&99&101&-0.19&-0.23\\
5618.633&495&45& \ldots&0.25& \ldots&6232.641&84&56&55&-0.31&-0.38\\
5620.493&43&35&\ldots&0.19& \ldots&  6254.258&123&95& \ldots&-0.04& \ldots\\
5633.946&71&43&37&-0.30&-0.43&6265.134&84&72&57&0.11&-0.20\\
5635.822&35& \ldots&17& \ldots&-0.11&6270.223&51& \ldots&33& \ldots&0.03\\
5638.262&78&45& \ldots&-0.31& \ldots&6301.501&122& \ldots&94&\ldots&-0.22\\
5679.024&60&51& \ldots&0.13& \ldots& 6302.494&84&74& \ldots&-0.00 & \ldots\\
5693.630&46&55& \ldots&0.38& \ldots& 6322.685&75&47&47&-0.14&-0.20\\
5717.833&63&41& \ldots&-0.11& \ldots& 6336.824&108&83& \ldots&0.06& \ldots\\
5731.762&58&30& \ldots&-0.22& \ldots&6344.148&61&24& \ldots&-0.28& \ldots \\
5741.848&31&32&29&0.37&0.26& 6355.028&70&42&43&-0.14&-0.19\\
5752.032&55& \ldots&24& \ldots&-0.41 &6393.600&132&94&89&-0.13&-0.30 \\
5762.992&120&97&81&0.17&-0.16 & 6408.018&96&66& \ldots&-0.33& \ldots\\
5816.373&81& 70&62& 0.00&-0.19&6411.649&134&81&72&-0.22&-0.43\\
5859.586&73& \ldots&59& \ldots&0.01& 6593.870&85&50& \ldots&-0.25& \ldots\\
5862.357&87&72&70&-0.37&-0.45&6597.559&45& \ldots&36& \ldots&0.08 \\
5883.817&67& \ldots&52& \ldots&0.14& 6609.110&65&37&30&-0.10&-0.29\\
5934.655&79& \ldots&51& \ldots&-0.13& 6705.102&47&43&33&0.21&-0.03\\
5983.68&702& \ldots&41& \ldots&-0.36&6726.666&47&27&28&-0.11&-0.14 \\
5984.815&86&61& \ldots&0.18& \ldots& \ldots&\ldots&\ldots&\ldots&\ldots&\ldots\\
\enddata
\tablecomments{Iron abundances for both components of V505~Per were determined using model atmospheres with $T_{\rm eff}=6650 \pm 50 $~K for star~A and $T_{\rm eff}=6550 \pm 50$~K for star~B.}
\end{deluxetable*}

\renewcommand{\arraystretch}{1.5}
\begin{deluxetable*}{cccccccc}
\tablecaption{Other abundances for both components of the binary system V505 Per\label{tab1}}
\tabletypesize{\footnotesize}        
\tablehead{
\colhead{WL (\AA)} & \colhead{Species} & \colhead{$\log gf$} & \colhead{$EW_{SUN}$} &
\colhead{EW$_A$ (m\AA)} & \colhead{EW$_B$ (m\AA)} &
\colhead{[X/H]$_A$} & \colhead{[X/H]$_B$}
}
\startdata
5688.205 & Na& -0.45 & 125& 87    & \ldots  & 0.11 & \ldots \\
5675.417 & Si& -1.03 &69& 48  &  52  & -0.12 & -0.09\\
5690.425 & Si&-1.87 & 49& 56 &32 & 0.18 & -0.26\\
5948.541 & Si&-1.23 &90& \ldots  & 68  & \ldots & 0.57 \\
6145.016 & Si&-0.82 &38& \ldots  & 65   & \ldots  & 0.43 \\
6155.134 & Si&-0.8 &88 & \ldots &77&-0.02\\
6237.319 & Si&-0.53&64 & 75  & 52  & 0.30 & -0.05 \\
5581.965 & Ca&-0.71 &95& 78  & \ldots    & -0.10 & 6.28\\
6166.439 & Ca&-0.90&70& 53 & \ldots  & 0.11& \ldots\\
6471.662 & Ca&-0.59 &92 & 52 & \ldots    & -0.41 & \ldots \\
6499.650 & Ca&-0.59&65 & 50  & \ldots  & -0.32 & 5.97 \\
6013.513 & Mn&-0.25&86&48  & \ldots & -0.06 & \ldots \\
6016.673 & Mn&-0.22&106 & 46  & 46  &-0.35 & -0.39\\
5754.656 & Ni& -2.22 &76& 28 &\ldots  & -0.55 & \ldots \\
6108.116 & Ni&-2.60 &64& 46 & 52     & 0.06  & 0.09 \\ 
6175.360 & Ni&-0.53 &50& 50 &\ldots   & 0.28  & \ldots \\
6223.984 & Ni&-0.91&28 & 15 & 14     & 0.00 & 0.13 \\
6643.630 & Ni&-2.22 &95& 50 & \ldots    & -0.45 & \ldots \\
6767.772 & Ni&-2.14 &79& \ldots & 39   & \ldots & -0.52\\
\enddata
\end{deluxetable*}

\floattable
\setlength{\tabcolsep}{10pt}       
\renewcommand{\arraystretch}{1.6}   
\begin{deluxetable*}{lcccc}
\tablecaption{Elemental abundances [X/H] and [X/Fe] for V505~Per stars A and B\label{tab2}}

\tablehead{
\colhead{Species} & \colhead{[X/H]$_A$} & \colhead{[X/H]$_B$} & \colhead{[X/Fe]$_A$} & \colhead{[X/Fe]$_B$}
}
\startdata
Fe &$-0.10 \pm 0.06$  & $-0.19 \pm 0.07$ & \ldots & \ldots\\
Na & $+0.11 \pm 0.08$ & \ldots & $+0.21 \pm 0.01$ & \ldots\\
Si &$+0.12 \pm 0.12$  &$+0.10 \pm 0.24$& $+0.22 \pm 0.13$  &$+0.29 \pm 0.25$\\
Ca &$-0.18 \pm 0.13$  &$-0.19 \pm 0.06$ &$-0.08 \pm 0.14$  &$0.00 \pm 0.09$\\
Mn &$-0.18 \pm 0.07$  & $-0.37 \pm 0.06 $ &$-0.07 \pm 0.09$  & $-0.18 \pm 0.09 $\\
Ni &$-0.13 \pm 0.17$  & $-0.10 \pm 0.21$ &$-0.03 \pm 0.18$  & $-0.09 \pm 0.22$ \\
\enddata
\end{deluxetable*}

\vspace{-2cm}
\section{Spectral Analysis and Abundance Determination}

In this section, we determine the chemical abundances of both components of V505~Per using an equivalent width analysis. The iron abundance in V505 Per was determined from equivalent width measurements using the  MOOG  spectrum analysis code \citep{1973PhDT.......180S} in conjunction with MARCS model atmospheres \citep{2008A&A...486..951G}. 

Abundances were computed using the assumption of local thermodynamic equilibrium (LTE). The adopted stellar parameters—effective temperature ($T_{\rm eff}$), surface gravity $(\log g)$, and micro-turbulent velocity—were derived as described in Section 4.

Stellar and solar abundances were calculated for each spectral line. Differential abundances for each component of V505 Per were then computed by subtracting the solar values from the stellar abundances in a line-by-line analysis, giving abundances relative to the solar scale. For the solar model atmosphere, we adopted $T_{\rm eff}=5780$ K and $\log g=4.4$ from the MARCS grid \citep{2008A&A...486..951G}. This approach is independent of the \(\log gf\) value for each line.
\vspace{1cm}

\subsection{Iron Abundance Fe I }

The equivalent width and corresponding abundance for each Fe line used in the analysis are listed in Table~\ref{tab:FeI}. Our analysis yields [Fe/H] = $-0.10 \pm 0.06$ for component A and [Fe/H] = $-0.19 \pm 0.07$ for component B. The difference in [Fe/H] between the two components falls within the quoted uncertainties. Therefore, we consider the two stars to have comparable metallicities. The determination of the uncertainties is described in section 6.

\subsection{Lithium Abundance Li I }
The lithium abundance was derived from the Li~\textsc{I} $\lambda$6707.76\,\AA ~ line. The measured equivalent widths are $26.6$\,m\AA\ for component A and $21.4$\,m\AA\ for component B. Since the Li~\textsc{I}  feature is blended and includes hyperfine structure, we analyzed it using the blends driver in  MOOG, supplied with a full line list that incorporates the hyperfine components and nearby blended lines \citep{2011ApJS..194...35D}.

We determined the lithium abundance in Star~A to be  $\log\,\epsilon$ (Li)~=~ 2.57 $\pm 0.09$ at an effective temperature of $6650 \pm 50$\,K, and in Star~B to be  $\log\,\epsilon$ (Li)~=~ 2.47 $\pm 0.12$ at $6550 \pm 50$\,K. We note that \citet{2025arXiv250719954K} employed spectrum synthesis in their analysis (see their Figure 6), and find that our results are consistent with theirs.

Since only a single Li line was available, to evaluate the internal scatter, we performed four independent measurements of the equivalent width by varying the placement of the continuum within reasonable bounds.

However, systematic uncertainties arising from stellar parameters—such as effective temperature, surface gravity, and microturbulence—as well as continuum placement, contribute to the total uncertainty. These are estimated to be ±0.09~dex for the primary star and ±0.12~dex for the secondary as discussed in section 6.\\

\begin{deluxetable*}{lccccc}
\tablecaption{Uncertainties in Abundance Determinations \label{tab:uncertainties}}
\tablehead{
\colhead{Element} & \colhead{Measurement } & \colhead{$\Delta T_{\rm eff}$} &
\colhead{$\Delta \log g $} & \colhead{$\Delta\xi_{\mathrm{turb}}$}& \colhead{Total Error}
}
\startdata
$\log\,\epsilon (Li)$  & 0.08 & 0.03 & 0.00 & 0.001 & 0.09 \\
$[Na/H]$& 0.06 & 0.03& -0.02 & 0.04 & 0.08 \\
$[Si/H] $ & 0.12 & 0.02 & -0.01 & 0.02 & 0.12 \\
$[Ca/H]$  & 0.12 & 0.03 & -0.01& 0.03 & 0.13 \\
$[Mn/H]$  & 0.04 & 0.02 & 0.00 & 0.05 & 0.07 \\
$[Fe/H] $ & 0.03 & 0.03 & -0.01 & 0.04 & 0.06 \\
$[Ni/H] $ & 0.16 & 0.03 & 0.00 & 0.04 & 0.17 \\
\enddata
\tablecomments{Measurement errors are SEM for Fe, Si, Ca, and Ni, and continuum-placement ranges for Li, Na, Mn, and Ba. Systematic uncertainties correspond to abundance changes for $\Delta T_{\rm eff}=\pm50$ K, $\Delta\log g=\pm0.1$, and $\Delta\xi_{\mathrm{turb}}=\pm 0.2 ~km s^{-1}$. The total uncertainty is the quadrature sum of the components.}
\end{deluxetable*}
\begin{deluxetable*}{ccccc}
\tablecaption{Comparison to Published Results \label{tab3}}
\tablehead{
\colhead{Literature} & \colhead{$T_{\mathrm{eff}}$(A)} & \colhead{$T_{\mathrm{eff}}$(B)} & \colhead{[Fe/H]$_A$} & \colhead{[Fe/H]$_B$}
}
\startdata
Tomasella et al. (2008)& $6512 \pm 50$ & $6460 \pm 50$& $-0.12 \pm 0.03$ & $-0.12 \pm 0.03$ \\
Baugh et al. (2013)& $6512 \pm 50$ & $6460 \pm 50$& $-0.15 \pm 0.03$ & $\ldots$ \\
Koenigsberger et al. (2025)& $6600 \pm 50$ & $6550 \pm 50$& $-0.15 \pm 0.05$ & $-0.15 \pm 0.05$ \\
This study  &$6650 \pm 50$ & $6550 \pm 50$ & $-0.10 \pm 0.06$ & $-0.19 \pm 0.07$\\
\enddata
\end{deluxetable*}
\vspace{-2cm}
\subsection{Other Elements}
In addition to iron and lithium, we determined the abundances of Na, Si, Ca,  Mn, and Ni from equivalent width measurements of carefully selected absorption lines in both components of V505 Per. The equivalent widths and derived abundances for each individual transition are listed in Table~\ref{tab1}, while the mean abundance for both components are summarized in Table~\ref{tab2}.

From two Si~I lines, we derived [Si/H] = $+0.12 \pm 0.12$ for star A and $+0.10 \pm 0.24$ for star B.

Several Na~I lines were initially considered; however, most were either too strong or blended to yield reliable results. We therefore focused on a single, unblended Na~I feature at 5688 $\AA$  for component A, which yielded [Na/H] = $+0.11 \pm 0.08$. The corresponding line in component B is blended and was not used for abundance determination. 

We identified many strong Ca~I absorption lines in the spectrum, but these could not be reliably analyzed because their large equivalent widths place them in the saturated part of the curve of growth, where abundance determinations are less reliable. 

For calcium, we obtained [Ca/H] = $-0.18\pm 0.13$ for component A and $-0.19 \pm 0.06$ for star B based on four weaker lines of Ca~\textsc{I}. Although the individual line measurements show some scatter, the mean values are consistent with the observed iron abundance. 

For nickel, we measured abundances using six lines in star A and 3 lines for star B. We derived [Ni/H] = $-0.13 \pm 0.17$ for star A and $-0.10 \pm 0.21$ for star B. 

To determine manganese abundances, which are strongly affected by hyperfine structure, we treated the Mn~\textsc{i} lines as blended features, following the approach of \citet{2011ApJS..194...35D}. For component A, we derived an abundance of [Mn/H] = $-0.18 \pm 0.07$, while for component B, we obtained [Mn/H] = $-0.37 \pm 0.06 $.

This generally represents the first determination of several elemental abundances for V505 Per.

\section{uncertainties}

A realistic evaluation of uncertainties is essential for interpreting the elemental abundances in V505~Per. We accounted for both random measurement errors and systematic uncertainties associated with the adopted stellar parameters.

For elements with multiple measurable transitions (e.g., Fe, Si, Ca, Ni), random errors are represented by the standard error of the mean (SEM) derived from line-to-line scatter.

For elements with only one or two detectable features (e.g., Li, Na, Mn ), random errors were estimated based on the range of abundances obtained using different continuum placements, which also captures the effect of local continuum setting on equivalent-width measurements.

Systematic uncertainties were assessed by perturbing the stellar parameters within their adopted uncertainties of $\pm50$~K in effective temperature ($T_{\mathrm{eff}}$), $\pm0.1$~dex in surface gravity ($\log g$), and $\pm0.2$~km~s$^{-1}$ in microturbulent velocity \(\xi_{\mathrm{turb}}\)-and repeating the abundance analysis.

The total uncertainty for each element was calculated as the quadrature sum of the measurement error and the sensitivities to $T_{\mathrm{eff}}$, $\log g$, and \(\xi_{\mathrm{turb}}\) (Table~\ref{tab:uncertainties}).

In most cases, uncertainties in the equivalent widths are the largest source of error. Uncertainties in microturbulence also contribute significantly in some species, but the resultant abundances are not particularly sensitive to temperature or surface gravity.

\section{Discussion}

We begin by examining the effective temperatures adopted for the two components of V505\,Per. We determine $T_{\rm {eff}} = 6650 \pm 50$~K for star A and $6550 \pm 50$~K for star B. These values are systematically higher than those reported by \citet{2008A&A...480..465T, 2013PASP..125..753B, 2021Obs...141..234S}, but are in close agreement with the temperatures derived by \citet{2025arXiv250719954K}.  

Importantly, \citep{2021Obs...141..234S} constrained the temperature difference between the two components of V505~Per to just 50\,K—a result that is fully consistent with our findings within the stated uncertainties. 

From the individual values of [Fe/H] = $-0.10 \pm 0.06$ for star A and [Fe/H] = $-0.19 \pm 0.07$ for star B using effective temperatures of $ 6650 \pm 50$ ~K for A and $6550 \pm 50$ ~K for B, we derive an average metallicity for the V505~Per system of [Fe/H] = $-0.15 \pm 0.05$, where the quoted uncertainty represents the root-mean-square (rms) error of the two components.

Previous studies have reported similar system-wide metallicities for V505 Per, with [Fe/H] values of $-0.12$, $-0.15$, and $-0.15$ from \citet{2008A&A...480..465T, 2013PASP..125..753B, 2025arXiv250719954K}, as summarized in Table~\ref{tab3}.

Overall, the derived abundances are consistent with those of Population I thin-disk stars, indicating that V505 Per has a chemical pattern typical of normal Milky Way thin-disk stars rather than any obvious abundance peculiarity \citet{adibekyan2012chemical}.

[Ca/Fe] and [Ni/Fe] are consistent with solar values within our uncertainties, while [Na/Fe] and [Si/Fe] show mild enhancements of about +0.2 dex (Table~\ref{tab2}). These abundance ratios are typical of thin-disk stars at similar metallicity \citep{adibekyan2012chemical}.

Manganese abundance ratios in Galactic disk stars are known to vary systematically with metallicity (e.g., \citealt{nissen2000structural}). In this context, the Mn abundances derived for the two components of V505 Per are mildly subsolar at [Mn/Fe]= -0.07 for star A and -0.18 for star B and remain compatible with the behavior commonly observed among thin-disk stars at similar metallicity \citep{sobeck2006manganese, mishenina2015new, adibekyan2012chemical}.

We investigated the lithium abundances of both components and found that the derived values are consistent with their effective temperatures and positions near the hot edge of the Li dip. Our results are in good agreement with those of \citet{2025arXiv250719954K}, who reported lithium abundances of $2.65 \pm 0.07$ and $2.35 \pm 0.10$ for stars A and B, respectively, at temperatures of 6600~K and 6550~K. This result supports the suggestion by \citet{2013PASP..125..753B} that the apparent inconsistency in lithium abundance may arise from the adopted temperature scale.

Taken together, the metallicity and element-ratio pattern of V505 Per reveal a system that is chemically typical for its metallicity.

\section{Conclusion}

We have determined the effective temperatures and elemental abundances for both components of the eclipsing binary system V505~Per. Our analysis of more than 50 iron lines per star yields effective temperatures of $6650 \pm 50$~K for component A and $6550 \pm 50$~K for component B. At these temperatures, we derive [Fe/H] $= -0.10 \pm 0.06$ for star A and $-0.19 \pm 0.07$ for star B. The system-wide mean metallicity, [Fe/H] $= -0.15 \pm 0.05$, indicates that V505~Per is slightly metal-deficient and is in excellent agreement with previous studies.

The abundances of other elements are broadly consistent with this mild iron deficiency. Most species, including Na, Si, Ca, and Ni, exhibit abundances that closely follow the iron content.

Our measurements support the conclusion of \citet{2025arXiv250719954K} that both stars lie on the hot side of the Li dip based on their effective 
temperatures.
\section*{Acknowledgments}

We are grateful to an anonymous referee for a thoughtful review of this manuscript, which substantially improved the paper. Pilachowski is grateful for the support of the Daniel Kirkwood Endowment at Indiana University that made this research possible.

\nocite{*}

\bibliography{sample701}

\end{document}